\documentclass[aps,floats, superscriptaddress, amsmath,epsf,twocolumn, prl]{revtex4}
\usepackage{amsfonts, relsize, color}
\usepackage{graphics}
\usepackage{graphicx}
\usepackage{subfigure}
\usepackage{hyperref}

\begin{document}

\title{ Interacting line-node semimetal: Proximity effect and spontaneous symmetry breaking}

\author{Bitan Roy}
\affiliation{Condensed Matter Theory Center and Joint Quantum Institute, University of Maryland, College Park, Maryland 20742-4111, USA}
\affiliation{Department of Physics and Astronomy, Rice University, Houston, Texas 77005, USA}

\date{\today}

\begin{abstract}
Effects of short-range electronic interactions in a three-dimensional line-node semimetal that supports linearly dispersing quasiparticles around an isolated loop in the Brillouin zone are discussed. Due to vanishing density of states ($\varrho(E) \sim |E|$) various orderings in the bulk of the system, such as the antiferromagnet and charge-density-wave, set in for sufficiently strong onsite ($U$) and nearest-neighbor ($V$) repulsions, respectively. While onset of these two orderings from the semimetallic phase takes place through continuous quantum phase transitions, a first-order transition separates two ordered phases. By contrast, topologically protected drumhead shaped surface states can undergo charge or spin orderings, depending on relative strength of $U$ and $V$, even when they are sufficiently weak. Such surface orderings as well as weak long-range Coulomb interaction can be conducive to spontaneous symmetry breaking in the bulk for weaker interactions. We numerically establish such proximity effect driven spontaneous symmetry breaking in the bulk for subcritical strength of interactions due to flat surface band and also discuss possible superconducting orders in this system.
\end{abstract}

\maketitle

\vspace{10pt}

\emph{Introduction}: The mechanism of mass generation of elementary particles through spontaneous symmetry breaking nowadays leaving the territory of high-energy physics curves its path through the fertile ground of condensed matter systems, where a myriad of gapless phases  emerges from complex band structures in solids. A celebrated example of mass generation is the \emph{superconductivity}, through which gapless excitations residing in the vicinity of the Fermi surface acquire \emph{Majorana mass}~\cite{mass-definition}.

Often the energy landscape available for electrons, offered by the periodic potential accommodated by immobile ions, displays band touching only at high symmetry points in the Brillouin zone; giving rise to semimetallic phase when the chemical potential is pinned at the band touching point~\cite{herring, dornhaus}. Although such band touching is usually protected by underlying symmetries, a plethora of broken symmetry phases (BSPs), lacking discrete and/or continuous symmetries, can be realized when electronic interactions are taken into account. In particular, we here focus on a three-dimensional system that supports linearly dispersing gapless excitations around an isolated two-dimensional \emph{loop} in the reciprocal space, the \emph{line-node semimetal} (LNSM), and address the effects of short-range electronic interactions on such unconventional phase of matter. Recently, developing interest in LNSM~\cite{ching-kai, sigrist, kim-NFL, kim-collectivemode, yan-collectivemode, nevidomskyy, nandkishore,moon-instability,lih-king}, along with its possible realization in various materials~\cite{PbTaSe-1, CaP, PbTaSe-2, TlTaSe, balents, LnIO, aji, carbon-1, carbon-2, uchoa, CuN, CuPdN, hyart-1, CaAgX, LaX, phos, murakami, goswami-roy, vanderbilt, ando, volovik-1}, where the strength of electronic interactions varies over a wide range, besides fundamental importance also endows timeliness to this quest.

\begin{figure}[htb]
\subfigure[]{
\includegraphics[width=4.0cm,height=4.0cm]{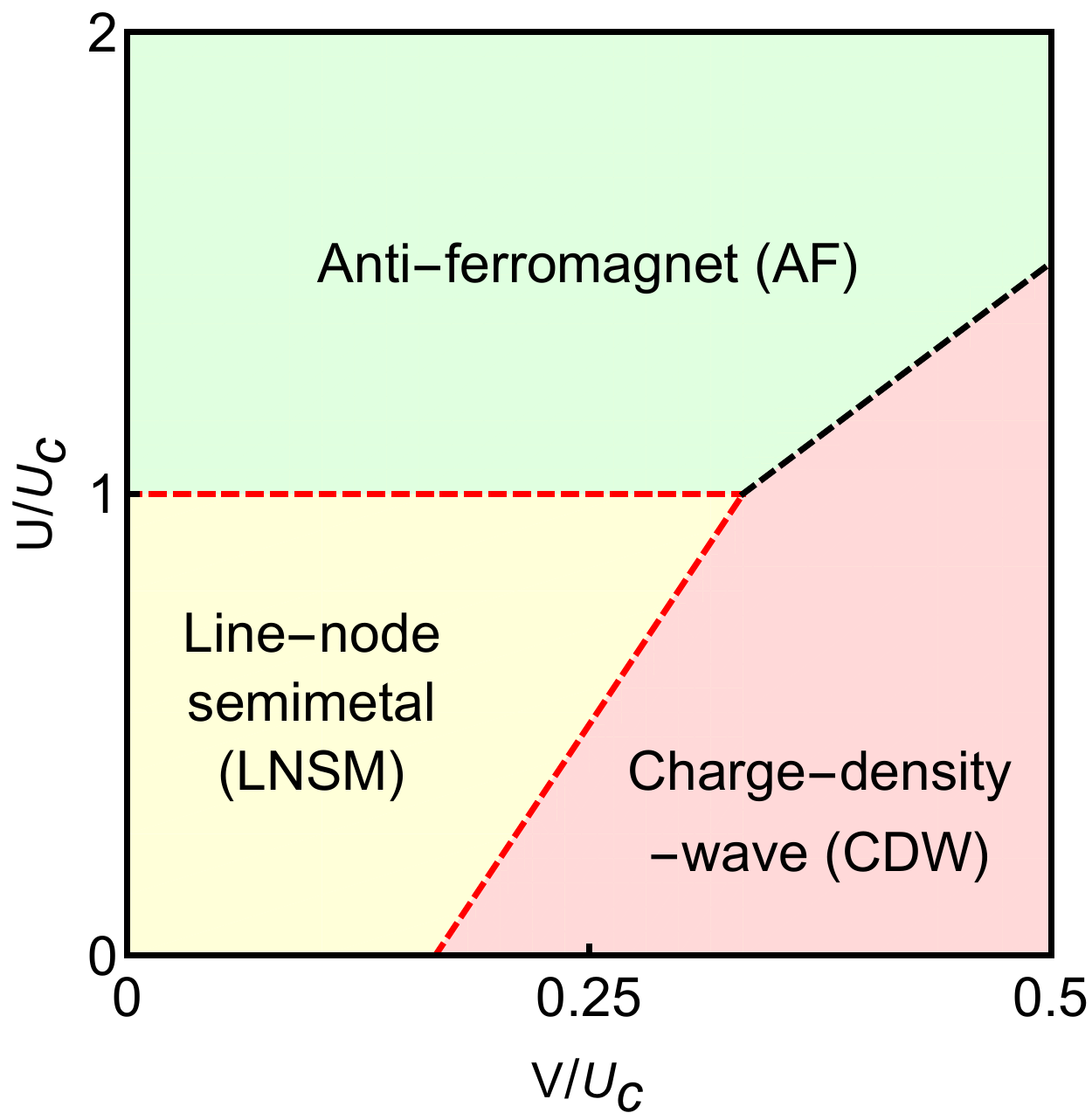}
\label{PhaseDiagram}
}
\subfigure[]{
\includegraphics[width=4.0cm,height=4.0cm]{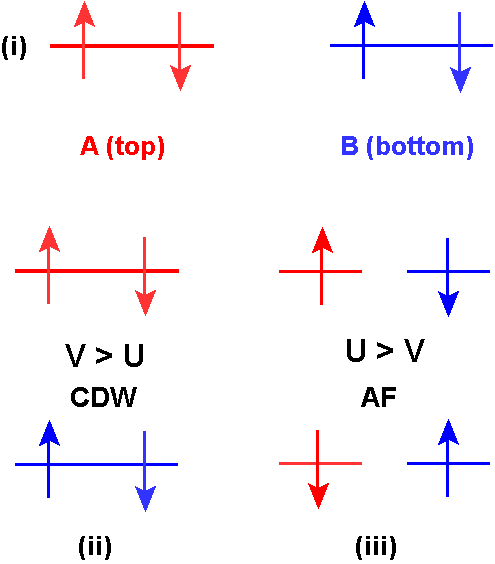}
\label{surfacesplitting}
}
\caption{(a) Mean-field phase diagram of a three-dimensional interacting LNSM. Here, $U$ and $V$ respectively correspond to onsite and nearest-neighbor repulsions [see Eq.~(\ref{extendedHubbard})], and support AF and CDW phases, when sufficiently strong. Critical strengths for AF ordering when $V=0$ is denoted by $U_c$. The transition(s) across the black (red) dashed line(s) is (are) first order (continuous). (b) (i) Spin degenerate surface states, localized on complimentary sublattices on opposite faces along the (001) direction. Two possible splittings for flat surface states: (ii) when $V>U$ and (iii) $U>V$. The surface ordering ordering sets in for infinitesimal strenght of $U$ and $V$, which in turn can trigger corresponding orderings in the bulk for weaker interactions, through the \emph{proximity effect}. When $U>V$, each surface is ferromagnet, but the magnetic moment points in opposite direction on opposite surfaces, and surface ordering assumes the form of layer or global AF (see also Fig.~\ref{self-consistentanalysis}). }\label{fig1:LNSM}
\end{figure}

\begin{figure}
\includegraphics[width=4.2cm, height=3.5cm]{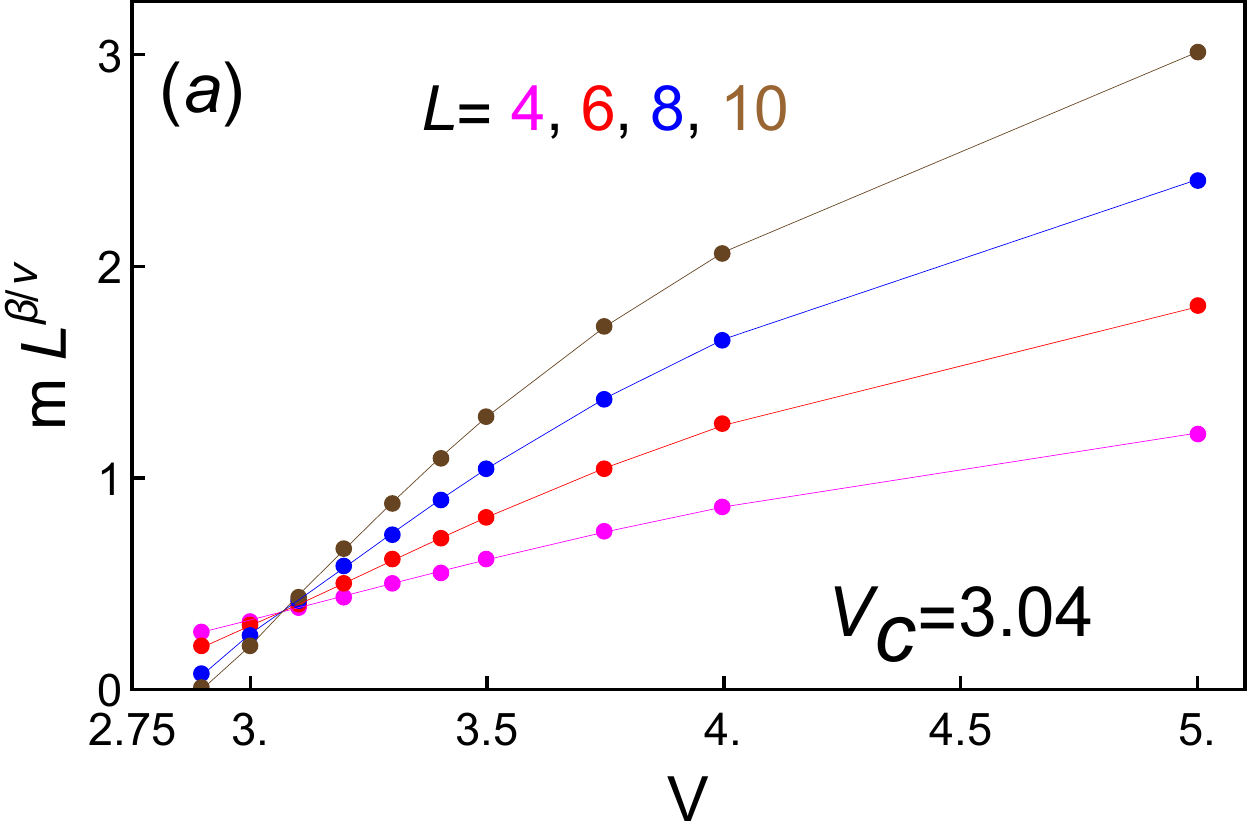}
\includegraphics[width=4.2cm, height=3.5cm]{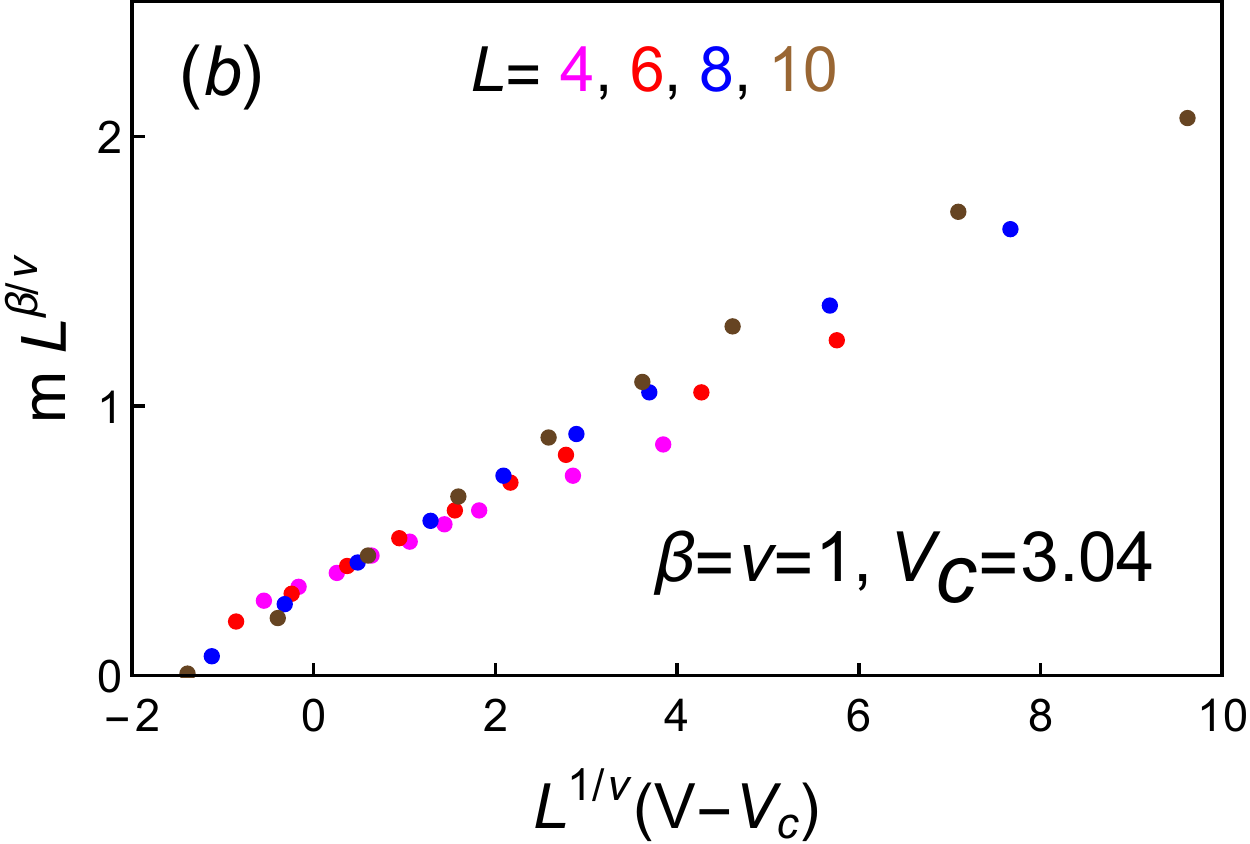}
\includegraphics[width=4.2cm, height=3.5cm]{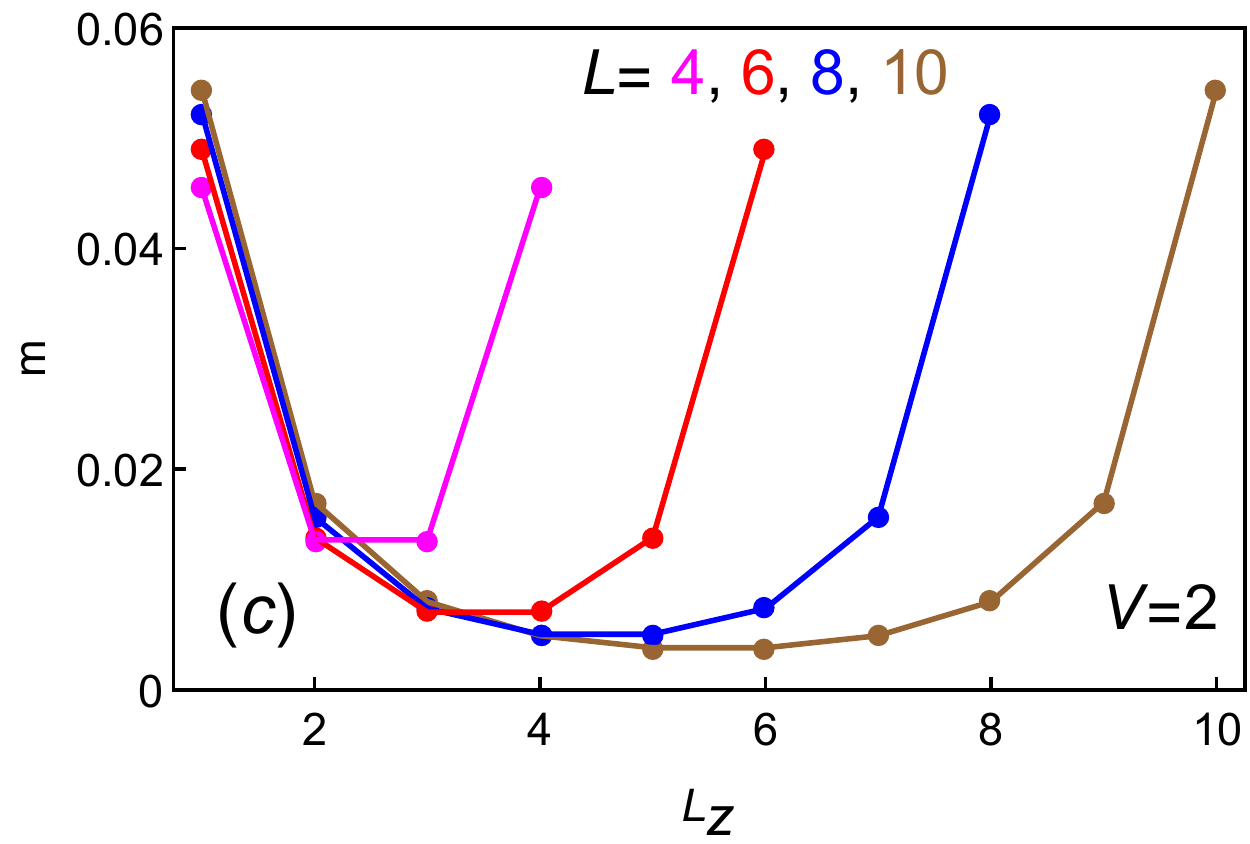}
\includegraphics[width=4.2cm, height=3.5cm]{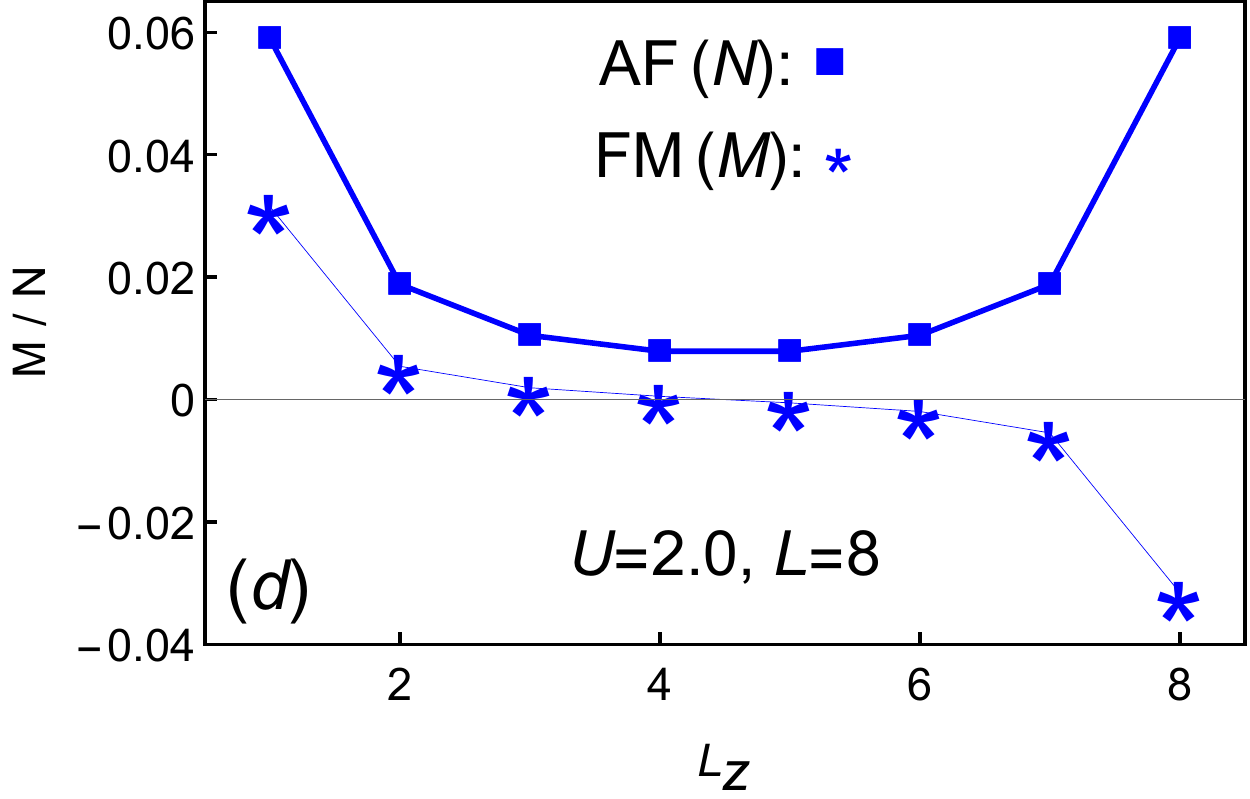}
\caption{ Hartree-Fock (mean-field) self-consistent solution of CDW [(a)-(c)] and AF/FM [(d)] orders with only nearest-neighbor ($V$) and onsite ($U$) repulsion, respectively [see Eq.~(\ref{extendedHubbard})]. Periodic boundary is always imposed along $x$ and $y$ direction, along which system size is $L_x=L_y=10$. Here $L$ represents the system in $z$ direction, along which we implement periodic boundary for (a) and (b) only (thus devoid of drumhead surface states). Throughout we set $t_{1,2,3}=t=1$, $b=1$ [see Eq.~(\ref{tightbinding})] and here $m=\Delta_{03}/t$, $M=\Delta_{j0}/t$, $N=\Delta_{j3}/t$. $U, V$ are measured in units of $t$. (a) CDW OPs in various systems ($L$) cross at a fixed value of $V$, yielding a critical strength of NN interaction $V_c=3.04$ for CDW ordering in the absence of surface states. (b) Data collapse shows that bulk undergoes a continuous transition at $V=V_c$. We here take the OP exponent $\beta=1$ and $\nu=1$ (large-$N$ values, see text) following a general scaling argument~\cite{supplementary}. (c) Spatial variation of CDW OP along $z$ direction for sub-critical strength of NN interaction ($V<V_c$), supporting the proposed proximity-induced ordering in the bulk due to surface states (see text). $L_z$ in panel (c) and (d) is the layer index along the $z$ direction. (d) Variation of FM ($M$) and AF ($N$) orders along the $z$ direction for weak onsite repulsion ($U=2$), showing that the ferromagnetic moment on two opposite surfaces points in opposite direction, yielding net zero magnetization. But, AF possesses same sign in the entire system, supporting a \emph{global anit-ferromagnet order}. When we impose periodic boundary, in all three directions $M=N=0$ everywhere in the system for $U=2$, suggesting ordering [in panel (d)] is purely due to the proximity effect.  }~\label{self-consistentanalysis}
\end{figure}

Due to \emph{linearly} vanishing density of states (DOS)~\cite{definition}, any sufficiently weak local four-fermion interaction (Hubbard-like) is an \emph{irrelevant} perturbation in a three dimensional LNSM. However, beyond a critical strength of interaction various BSPs can set in through continuous quantum phase transitions (QPTs). We here identify the antiferromagnet (AF) and charge-density-wave (CDW) as favorable BSPs that, for example, can be realized for sufficiently strong onsite Hubbard ($U$) and nearest-neighbor ($V$) repulsions, respectively; see Fig.~\ref{PhaseDiagram}. The LNSM also supports topologically protected flat \emph{drumhead} shaped surface states, which are susceptible to either charge or spin orderings even for infinitesimal interaction; see Fig.~\ref{surfacesplitting}. In turn the surface orderings can induce aforementioned orders in the bulk through \emph{proximity effect} even at weaker interactions, which we numerically demonstrate through a Hartree-Fock self-consistent calculation in a finite system of LNSM (see Fig.~\ref{self-consistentanalysis}). Thus surface-induced ordering in the bulk stands as an experimentally feasible way to induce BSPs in the bulk, as well as drive the system through QPTs by systematically tuning the thickness of the sample~\cite{finitesize}.

\emph{Model}: A LNSM can be realized from the following simple tight-binding Hamiltonian on a cubic lattice
\begin{eqnarray}\label{tightbinding}
H &=& t_1 \sum_{\mathbf{k}} \Psi^\dagger_{\mathbf{k}} \left[ \cos(k_x a)+\cos(k_y a) -b  \right] \tau_1 \Psi_{\mathbf{k}} \nonumber \\
&+& \sum_{\mathbf{k}} \Psi^\dagger_{\mathbf{k}} \left[ t_2 (\cos(k_z a)-1) \tau_1 + t_3 \sin(k_za) \tau_2 \right] \Psi_{\mathbf{k}},
\end{eqnarray}
where $a$ is the lattice spacing. The two-component spinor is defined as $\Psi^\top_{\mathbf{k}}=\left( c_{A,\mathbf{k}}, c_{B,\mathbf{k}} \right)$. Here, $c_{j,\mathbf{k}}$ is the fermion annihilation operator on sublattice $j=A,B$ with momentum $\mathbf{k}$, and $\tau_1$ and $\tau_2$ are two off-diagonal Pauli matrices. The Hamiltonian is invariant under the reversal of time, generated by ${\mathcal T}=K$, where $K$ is the complex conjugation and ${\mathcal T}^2=+1$. The anticommutation relation $\left\{H, \tau_3 \right\}=0$ ensures the spectral symmetry of the system. The inversion symmetry ($\mathcal P$), under which $\mathbf{k} \to -\mathbf{k}$ and $\Psi_{\mathbf{k}} \to \tau_1 \Psi_{-\mathbf{k}}$ guarantees that an isolated nodal ring is pinned at the $k_z=0$ plane. A LNSM supports topologically protected flat surface state at zero energy (image of the bulk loop)~\cite{ching-kai, volovik-1, balents, nevidomskyy}, which for the above model is an eigenstate of $\tau_3$ with eigenvalue $\pm 1$. Therefore, the drumhead shaped surface states are localized on \emph{complimentary} subalattices on \emph{opposite} surfaces along the $(001)$ direction, see Fig.~\ref{surfacesplitting}(i).

Restoring the spin degrees of freedom and upon linearizing the above model near $\mathbf{k}=(0,0,0)$ point, we arrive at the continuum description of the LNSM
\begin{equation}
\hat{H}_{0}= \Gamma_1 \; \left( \frac{k^2_\perp-k^2_F}{2m} \right) + \Gamma_2 \; (v_z k_z) \equiv \sum_{j=1,2}\Gamma_j d_j,
\end{equation}  
where $m^{-1}=t_1 a^2$, $v_z=t_3 a$, $k^2_\perp=k^2_x+k^2_y$ and $k_F=\sqrt{2(2-b)} a^{-1}$, and $\Gamma_j=\sigma_0 \otimes \tau_j$. Pauli matrices $\sigma_\mu$ operate on the spin index. The radius of the nodal ring is $k_F$, around which $(k^2_\perp-k^2_F)/(2m) \approx v_r k_r$, where $v_r=k_F/m$ is the radial Fermi velocity. Thus fermionic dispersion scales \emph{linearly} with the radial ($k_r=k_\perp -k_F$) and $\hat{z}$ components of momentum, and a LNSM corresponds to a $z=1$ fixed point, where $z$ is the \emph{dynamic scaling exponent}. Besides $\mathcal{P}$ and $\mathcal{T} (=i\sigma_2 \otimes \tau_0 \; K)$ symmetries, $\hat{H}_{0}$ is also invariant under $SU(2)$ chiral rotation of the spin quantization axis, generated by $\boldsymbol \sigma \otimes \tau_0$.

\begin{table}[h]
\begin{tabular}{|c|c|c|c|c|c|}
\hline
{\bf OP} & {\bf \emph{Physical order}} & $\mathcal T$ & $\mathcal P$ & $SU(2)$ & {\bf \emph{Susceptibility}} \\
\hline \hline
$\Delta_{01}$ & Bond density & $\checkmark$ & $\checkmark$ & $\checkmark$ & $f(v_r, v_z) \Lambda/2$ \\
\hline
$\Delta_{02}$ & Current density & $\times$ & $\times$ & $\checkmark$ & $f(v_r, v_z) \Lambda/2$ \\
\hline
$\Delta_{03}$ & {\bf Charge-density-wave} & $\checkmark$ & $\times$ & $\checkmark$ & $f(v_r, v_z) \Lambda$ \\
\hline \hline
$\Delta_{j0}$ & Ferromagnet & $\times$ & $\checkmark$ & $\times$ & $0$ \\
\hline
$\Delta_{j1}$ & Spin bond density & $\times$ & $\checkmark$ & $\times$ & $f(v_r, v_z) \Lambda/2$ \\
\hline
$\Delta_{j2}$ & Spin current density & $\checkmark$ & $\times$ & $\times$ & $f(v_r, v_z) \Lambda/2$ \\
\hline
$\Delta_{j3}$ & {\bf Antiferromagnet} & $\times$ & $\times$ & $\times$ & $f(v_r, v_z) \Lambda$ \\
\hline
\end{tabular}
\caption{Various OPs, and their physical realizations and transformations under various symmetry operations. Here $\checkmark$ and $\times$ stand for even and odd under a symmetry operation, respectively. The last column displays mean-field susceptibility of various orderings, where $\Lambda$ is the ultraviolet cutoff and $f(v_r, v_z)$ is a nonuniversal, but positive definite function~\cite{supplementary}. }\label{tab-1}
\end{table}

\emph{Broken symmetry phases}: Due to vanishing DOS ($\varrho(E) \sim |E|$), a LNSM remains stable against sufficiently weak, but generic short-range, such as onsite Hubbard ($U$), nearest-neighbor ($V$) interactions. However, strong repulsive interactions can destroy the noninteracting $z=1$ fixed point and give rise to various BSPs. All together, a LNSM is susceptible to \emph{seven} types of \emph{intra-unit cell} excitonic instabilities. The corresponding effective single particle Hamiltonian reads as $H_{SP}=\Delta_{\mu \nu} \left( \sigma_\mu \otimes \tau_\nu \right)$, where $\mu,\nu=0,1,2,3$ (except for $\mu=\nu=0$, since $\Delta_{00}$ is chemical potential) and $\Delta_{\mu \nu}=\langle \Psi^\dagger \sigma_\mu \otimes \tau_\nu \Psi \rangle$ is the order parameter (OP). The physical meaning of these OPs and their transformation under symmetry operations are shown in Table.~\ref{tab-1}. Due to spin rotational symmetry, $\Delta_{j \nu}=|\vec{\Delta}_\nu|$ for $j=1,2,3$ and any $\nu$.

In the presence of underlying bond ($\Delta_{01}$) and spin bond ($\Delta_{j1}$) density orders, the nearest-neighbor hopping amplitude respectively acquires spin independent or dependent modulation, and the ordered phases support two nodal rings in the $xy$ plane, with radii $k^{\pm}_\perp=[k^2_F \pm 2 m \Delta]^{1/2}$, for $\Delta=\Delta_{01}$ or $|\vec{\Delta}_1|$. By contrast, nucleation of $\Delta_{02}$ and $\Delta_{j2}$ respectively gives rise to current density and spin current density. These two ordered phases are also accompanied by two nodal rings, but they are of identical radius $k_\perp=k_F$ and placed at $k_z=\pm \Delta/v_z$ for $\Delta=\Delta_2$ or $|\vec{\Delta}_{2}|$. The CDW ($\Delta_3$) and AF ($|\vec{\Delta}_3|$) orders, respectively gives rise to staggered pattern of charge and spin among the nearest-neighbor sites of the cubic lattice, and concomitantly to a \emph{fully gapped spectrum}. Onset of a ferromagnetic ordering leads to the formation of compensated electron and hole doped line nodes for opposite spin projections, and thus susceptible to a subsequent BCS-like excitonic instability toward the formation of the AF order. But, the AF OP gets locked into the spin easy-plane, perpendicular to the ferromagnetic moment, and represents \emph{canted antiferromagnet}. While the nodal orderings give rise to drumhead shaped surface states on the (001) surface in the ordered phases (images of the bulk nodal rings), onset of mass orders (such as CDW, AF) splits the surface states and places them at finite energies, as shown in Fig.~\ref{surfacesplitting}.

Valuable insight into the propensity toward various orderings can be gained from the corresponding static mean-field susceptibilities ($\chi$) (see Table~\ref{tab-1}). Note that the critical strength of interaction for any ordering is \emph{inversely} proportional to $\chi$. The CDW and AF orders possess the largest susceptibility, and consequently require minimal strengths of interaction for nucleation. Since at $T=0$ optimal minimization of the free energy (no competition with entropy) naturally prefers fully gapped phases, generic short-range interaction supports either CDW or AF orders over gapless phases. Despite being accompanied by \emph{two} massless Goldstone mode (due to spontaneous breaking of $SU(2)$ spin rotational symmetry) the AF phase in a three-dimensional LNSM can exhibit a genuine finite temperature continuous phase transition, described by a three-component $\phi^4$ theory, while that for the CDW phase is captured by an Ising $\phi^4$ theory.

\emph{Proximity effect}: Since surface states, arising from the model shown Eq.~(\ref{tightbinding}), are completely flat, they can undergo various weak coupling instabilities, such as surface CDW when $V>U$ and AF when $U>V$ as shown in Fig.~\ref{surfacesplitting}, due to \emph{diverging} DOS, before the bulk acquires propensity toward any ordering. Such surface ordering through the proximity effect in turn can give rise to BSP at weaker couplings in the bulk, as we demonstrate from a self-consistent solution of these OPs in a finite LNSM with open boundaries along $(001)$ direction (thus supporting drumhead surface states), see Fig.~\ref{self-consistentanalysis}. Therefore, in experiment one can observe various BSPs by systematically reducing the thickness of the system, i.e., in a thin film of LNSM~\cite{finitesize}. In addition, such proximity effect should be more pronounced in materials with a larger radius of the nodal-loop ($k_F$), as the number of surface states increases with increasing $k_F$. Thus, the thickness of a LNSM as well as the radius of the nodal-loop [tunable by hydrostatic pressure, for example, controlling the hopping parameters $t_{1,2,3}$, see Eq.~(\ref{tightbinding})] can be two experimentally accessible \emph{nonthermal} tuning parameters to drive the system through a QPT and realize various BSPs. It is worth noticing that when $U>V$, each surface along (001) direction becomes \emph{ferromagnetic} due to the sublattice polarization of surface states, see Figs.~\ref{surfacesplitting} and \ref{self-consistentanalysis}(d). The magnetic moment, however, points in opposite direction on opposite surfaces, and the ordered phase represents a \emph{layer or global antiferromagnet} [see Fig.~\ref{self-consistentanalysis}(d)]. Predicted surface change (spin) ordering can be detected by STM (spin-resolved STM) measurements. Next, we investigate the onset of CDW and AF orders and their competition inside the bulk of a LNSM.

\emph{Phase diagram}: The stability of a LNSM (for weak interactions), the possible onset of CDW, AF phases, and the competition between these two ordered phases (for strong interactions) can be demonstrated from the following mean-field free energy density 
\begin{eqnarray}\label{free-gen}
F &=&\frac{\Delta^2_3}{2 g_C}+ \frac{|\vec{\Delta}_3|^2}{2 g_{AF}} -2 \sum_{\sigma=\pm} \int^{\prime} \frac{d^3 \mathbf{k}}{(2 \pi)^3} \; E_\sigma (\mathbf{k}),
\end{eqnarray}
where $E_\sigma (\mathbf{k})=\left[ d^2_1+d^2_2+\Delta^2_3+|\vec{\Delta}_3|^2+ 2 \sigma \Delta_3 |\vec{\Delta}_3| \right]^{1/2}$. The integral over momentum is restricted up to an ultraviolet cutoff $\Lambda$. In the presence of onsite ($U$) and nearest-neighbor ($V$) repulsions (extended Hubbard model), the interacting Hamiltonian reads as 
\begin{equation}~\label{extendedHubbard}
H_{I}= U \sum_{\vec{x}} n_{\uparrow}(\vec{x}) n_{\downarrow}(\vec{x}) +\frac{V}{2} \sum_{\vec{a}, j, \sigma, \sigma^\prime} n_{\sigma}(\vec{a}) n_{\sigma^\prime}(\vec{a}+\vec{b}_j)
\end{equation}
and we obtain $g_C = (6 V-U)/8$ and $g_{AF} = U/8$. Here, $n_\sigma(\vec{x})$ is the fermionic number at $\vec{x}$, with spin projection $\sigma=\uparrow, \downarrow$ and sites on A (B) sublattice are located at $\vec{a}$ ($\vec{a}+\vec{b}_j$ where $j=1, \cdots, 6$).

Let us first focus on a simpler situation with only CDW order, i.e., when $|\vec{\Delta}_3|=0$. Minimizing $F$ we then obtain the following gap equation 
\begin{equation}
g_C \; \int^{E_\Lambda}_0 d\varepsilon \; \frac{\varrho(\varepsilon)}{\sqrt{\varepsilon^2 + \Delta^2_3}}=1,
\end{equation}
where $E_\Lambda$ is the ultraviolet energy up to which the quasiparticle dispersion in a LNSM is linear. The above gap equation yields a nontrivial solution for order parameters 
\begin{equation}
\delta_C=1+m_C - \sqrt{1+m^2_C}
\end{equation}
where $m_C=\Delta_3/E_\Lambda$ is the dimensionless CDW OP, $\delta_C=\left[ (g^\ast_C)^{-1} - g^{-1}_C \right]E^{-1}_\Lambda$, only when the interaction is stronger than the critical one ($g^\ast_C$), i.e. $g_C > g^\ast_C$. In close proximity to the quantum critical point (QCP), located at $g_C=g^\ast_C$, $m_C \ll 1$, yielding $m_C \sim \delta_C$. Therefore, the general scaling relation $m_C \sim \delta^{\nu z}_C$, suggests that $\nu z=1$, where $\nu$ is the \emph{correlation length exponent}. A similar self-consistent solution can be found for the AF order upon setting $\Delta_3=0$ in Eq.~(\ref{free-gen}). Hence, CDW and AF orderings set in beyond a critical strength of interaction through a continuous QPT and the transition temperature for these two orderings scales as $T_c \sim |\delta|^{\nu z}$. By solving the coupled gap equations and simultaneously minimizing the free energy~\cite{supplementary}, we arrive at the phase diagram of an interacting LNSM, shown in Fig.~\ref{PhaseDiagram}. The transition between AF and CDW orders is \emph{first order} in nature~\cite{RG-large-N}.

\emph{Long range interaction}: We now comment on the effect of long range tail of the Coulomb interaction. The leading corrections to the \emph{anomalous dimension} due to long-range Coulomb interaction is largest for \emph{mass} orders (CDW and AF) and proportional to $\alpha_r F(\eta)/\Lambda$, where $\alpha_r=e^2/(4 \pi^2 v_r \epsilon_0)$ is the fine structure constant in the radial direction, $\eta=v_z/v_r$, $F(\eta)=\mbox{E}_k(1-\eta^2)+\mbox{E}_k(1-\eta^{-2})/\eta$ and $\mbox{E}_k$ is the elliptic function of first kind~\cite{supplementary}. Such enhancement of anomalous dimension indicates that long-range interaction boosts the propensity toward these two orderings in a LNSM. Consequently, the phase boundaries between LNSM-BSPs, shown in Fig.~\ref{PhaseDiagram}, shift toward a weaker strength of interactions (such as $U$ and $V$) in particular when the dielectric constant ($\epsilon_0$) is small. But, a complete analysis on the interplay of a proposed non-Fermi liquid~\cite{kim-NFL} and BSPs in the presence of both long- and short-range components Coulomb interactions demands a separate investigation~\cite{katsnelson}.

\emph{Superconductivity}: Finally, we shed light on possible paired states in a LNSM, when the electronic interaction acquires a strong attractive component. We now introduce a Nambu doubled spinor as $\Psi_N=\left(\Psi_p, \Psi_h \right)^\top$, where $\Psi_p=\left( \Psi_{p, \uparrow},\Psi_{p, \downarrow} \right)^\top$, $\Psi_h=\left( \Psi_{h, \downarrow},-\Psi_{h, \uparrow} \right)^\top$ and $\Psi^{\top}_{p,s}=(\Psi^\dagger_{A,s}, \Psi^\dagger_{B,s} )$, $\Psi^{\top}_{h,s}=\left(\Psi_{B,s}, \Psi_{A,s} \right)$ for $s=\uparrow/\downarrow$. In this basis $\vec{S}=\eta_0 \otimes \vec{\sigma} \otimes \tau_0$ are the three generators of electron spin~\cite{BR-classification} and the non-interacting Hamiltonian becomes $\hat{H}^{N}_{0}=\eta_3 \otimes \Gamma_1 d_1+ \eta_0 \otimes \Gamma_2 d_2$, where the Pauli matrices $\eta_\mu$ operate on the Nambu index. All together the LNSM permits \emph{four} local pairings, and the corresponding effective single-particle Hamiltonian reads as $H_{SC}= \left(\eta_1 \cos \phi + \eta_2 \sin \phi \right) \otimes H_{p}$, where $\phi$ is the superconducting phase and $H_{p}=\Delta_{s} \Gamma_1 + \Delta_{0} \Gamma_0 + \Delta_{2} \Gamma_2  + \Delta_t \; \vec{\sigma} \otimes \tau_3$. First three pairings are spin-singlet, while the last one is spin-triplet, among which only the spin singlet $s$-wave pairing ($\Delta_s$) stands as \emph{Mojorana mass}~\cite{mass-definition}. The remaining three pairings support gapless BdG quasiparticles around nodal rings inside the ordered phase, and thus are expected to be energetically inferior to the fully gapped $s$-wave pairing~\cite{supplementary}. It should be noted that due to the presence of flat surface states the $s$-wave pairing can take place on the surface in even for sufficiently weak attractive interaction, which can also be mediated by electron-phonon interaction~\cite{hyart-2}. Such surface superconductivity can in turn induce pairing among the bulk states through the proximity effect, in particular for a thin film of LNSM.

Upon casting \emph{six} matrices corresponding to the mass orders (AF, CDW, $s$-wave pairing) in Nambu representation, we can arrange them into two sets according to $\left\{ \eta_3 \otimes \vec{\sigma} \otimes \tau_3\right\} \: \mbox{and} \: \left\{ \eta_0 \otimes \sigma_0 \otimes \tau_3, \vec{\eta}_\perp \otimes \sigma_0 \otimes \tau_1 \right\}$. Together they constitute a $Cl(3) \times Cl(3)$ algebra, where $\vec{\eta}_\perp=(\eta_1, \eta_2)$~\cite{cl3}. Therefore, both strong onsite Hubbard repulsion and attraction can destabilize a LNSM, by respectively supporting an AF order or through simultaneous nucleation of a CDW and $s$-wave superconductor. Respectively the ordered phase breaks the spin and pseudo-spin $SU(2)$ symmetries. These two transitions thus belong to the same universality class, which can be demonstrated in numerical simulations. Such exact symmetry between CDW and $s$-wave pairing stems from the bipartite nature of the underlying cubic lattice, the absence of particle-hole asymmetry in the normal state and any other finite range component of Coulomb interaction. For example, weak repulsive (attractive) nearest-neighbor interaction lifts such degeneracy and prefers CDW (s-wave pairing).

\emph{Conclusions}: To conclude, we here show that an interacting LNSM can be susceptible to a plethora of BSPs in the bulk (for strong interaction) as well as on the surface (for weak interaction). The weak coupling instabilities of drumhead shaped flat surface states can in turn induce orderings in the bulk through the proximity effect even for a weaker interaction, making our proposals relevant in real materials~\cite{PbTaSe-1, CaP, PbTaSe-2, TlTaSe, balents, LnIO, aji, carbon-1, carbon-2, CuN, CuPdN, CaAgX, LaX, phos, murakami, goswami-roy, vanderbilt, ando}, in particular for a thin film of LNSM (see Fig.~\ref{self-consistentanalysis}). By contrast, if a LNSM lacks ${\mathcal T}$ symmetry or spin degeneracy~\cite{goswami-roy}, coined as Weyl-loop semimetal, the number of possible ordering channels is quite restricted, with \emph{ferromagnet} being the only possible mass order~\cite{supplementary}. Nevertheless, our conclusions regarding the fate of the QPT and surface-induced proximity effect into bulk remains unchanged in this system.

\emph{Acknowledgments}: This work was supported by ARO-Atomtronics-MURI, NSF-JQI-PFC and LPS-MPO-CMTC. The author is grateful to Victor Galitski for continued interest in this work and many stimulating discussions. B. R. also thanks C-K. Chiu, I. Herbut, R-J. Slager for discussions; P. Goswami and V. Juri\v ci\' c for critical reading of the manuscript, and Nordita, Center for Quantum Materials for hospitality.


\begin{thebibliography}{99}
\bibitem{definition} DOS scales as $\rho(\varepsilon) \sim |\varepsilon|^{-1+d_c/z}$, where $d_c$ is the co-dimension, which equals to the difference between the physical dimension and dimensionality of the hyperplane over which the energy gap vanishes. The co-dimension of a LNSM (Fermi surface) is $d_c=2 (1)$, yielding linearly vanishing (constant) DOS, since $z=1$ for linearly dispersing quasiparticles around the Fermi energy. For $d$-dimensional Dirac or Weyl semimetals, $d_c=d$, yielding a standard scaling of DOS, $\rho(\varepsilon) \sim |\varepsilon|^{-1+d}$.
\bibitem{finitesize} The finite size effect or hybridization induced gap for the surface states ($\sim 1/L$) is assumed to be $\ll U$ or $V$.
\bibitem{cl3} $Cl(3)$ is a set of three mutually anticommuting matrices.
\end{thebibliography}
\end{document}